\documentclass[sigconf]{acmart}

\usepackage{graphicx}
\graphicspath{ {./} }

\usepackage[skip=1ex]{caption}
\usepackage{ragged2e}
\usepackage{booktabs,tabularx}
\newcommand\mcc[1]{\multicolumn{2}{c}{#1}}

\AtBeginDocument{%
  \providecommand\BibTeX{{%
    \normalfont B\kern-0.5em{\scshape i\kern-0.25em b}\kern-0.8em\TeX}}}

\setcopyright{acmcopyright}
\copyrightyear{2023}
\acmYear{2023}
\acmDOI{10.1145/1122445.1122456}
\acmConference[ICPE2023]{ICPE 2023: International Conference on Performance Engineering}{April 15-19, 2023}{Coimbra, Portugal}
\begin{document}

\title{Hunter: Using Change Point Detection to Hunt for Performance Regressions}

\author{Matt Fleming}
\email{matt@codeblueprint.co.uk}
\authornote{DataStax, Inc}
\affiliation{%
  \institution{~}
}

\author{Piotr Kołaczkowski}
\email{pkolaczk@datastax.com}
\authornotemark[1]
\affiliation{%
  \institution{~}
}

\author{Ishita Kumar}
\email{ishitakumar@umass.edu}
\authornotemark[1]
\affiliation{%
  \institution{~}
}

\author{Shaunak Das}
\email{shaunak.das@datastax.com}
\authornotemark[1]
\affiliation{%
  \institution{~}
}

\author{Sean McCarthy}
\email{sean.mccarthy@datastax.com}
\authornotemark[1]
\affiliation{%
  \institution{~}
}

\author{Pushkala Pattabhiraman}
\email{Pushkala.Pattabhiraman@datastax.com}
\authornotemark[1]
\affiliation{%
  \institution{~}
}

\author{Henrik Ingo}
\email{henrik.ingo@avoinelama.fi}
\authornotemark[1]
\affiliation{%
  \institution{~}
}

\renewcommand{\shortauthors}{Fleming and Kołaczkowski, et al.}

\begin{abstract}
Change point detection has recently gained popularity as a method of detecting performance changes in software due to its ability to cope with noisy data. In this paper we present Hunter, an open source tool that automatically detects performance regressions and improvements in time-series data. Hunter uses a modified E-divisive means algorithm to identify statistically significant changes in normally-distributed performance metrics. We describe the changes we made to the E-divisive means algorithm along with their motivation. The main change we adopted was to replace the significance test using randomized permutations with a Student's t-test, as we discovered that the randomized approach did not produce deterministic results, at least not with a reasonable number of iterations. In addition we've made tweaks that allow us to find change points the original algorithm would not, such as two nearby changes. For evaluation, we developed a method to generate real timeseries, but with artificially injected changes in latency. We used these data sets to compare Hunter against two other well known algorithms, PELT and DYNP. Finally, we conclude with lessons we’ve learned supporting Hunter across teams with individual responsibility for the performance of their project.

\end{abstract}

%



\maketitle

\section{Introduction}
Testing the performance of distributed databases, such as Apache Casandra, is an integral part of
the development process and is often incorporated into Continuous Integration pipelines where
performance tests and benchmarks can be run periodically or in response to pushing changes to source
code repositories. But given the complex nature of distributed systems, their performance is often
unstable and performance test results can fluctuate from run to run even on the same hardware. This
result instability is due to a number of factors including variability of the underlying hardware
\cite{ONLYCONSTANTISCHANGE}, background processes and CPU frequency scaling at the OS level, and
application-level request scheduling and prioritisation \cite{TALESOFTHETAIL}. All of this makes the
job of identifying whether the change in performance is the result of a software change or simply
noise from the test extremely difficult to do automatically. Threshold-based techniques are covered
in the literature, but these methods do not handle noise in benchmark data well and require that
threshold values be set per-test \cite{MONGOCPD}. Additionally, thresholds need to be periodically 
tuned as performance improvements are added and new baselines are established.

In the past, we have relied heavily on experienced engineers to visually inspect graphs and benchmark
data to identify changes in performance. However, this suffers from a number of drawbacks including:

\begin{itemize}
\item Expert knowledge for identifying changes is difficult to teach other engineers
\item Small teams of experts have a limit on the number of tests they can inspect
\item Even experienced engineers can miss changes
\end{itemize}

Because of these drawbacks, we have recently created Hunter\cite{HUNTER}, an open source tool that uses change point detection to find statistically significant changes in time-series data. Change point detection has recently gained favour as a method of coping with the inherent instability, or noise, in performance test and benchmark data \cite{MONGOCPD} and can identify both performance regressions and improvements.

Hunter was designed with the goal of eliminating the need for a dedicated group of engineers to sift through performance test results. Instead, individual teams can feed their benchmark data to a central datastore which Hunter pulls from and analyses. We use Hunter for validating multiple releases across various distributed database and streaming products which has required that we make Hunter intuitive and user-friendly for engineers that are experts in their particular area but not performance experts.

The contributions in this paper are:

\begin{itemize}
\item We present an open-source tool that can run change point detection on any time-series data containing multiple metrics in either a .CSV file or stored on a graphite server.
\item We discuss the modifications we have made to the E-divisive means algorithm to improve its performance and predictability of results.
\item We develop a method for generating timeseries of real benchmarking results, with artificially injected changes to latency at discrete points in time. This allows us to evaluate the accuracy of an algorithm objectively, against a known set of correct change points.
\item We compare Hunter (modified E-divisive means) against two other change point detection algorithms, DYNP and PELT.
\item We share the lessons that we have learned from running Hunter in a multi-team environment where each team is responsible for a different product and favors different benchmarks.
\end{itemize}

\section{High-Level Overview}
Hunter is a command-line tool, written in Python, that detects statistically significant changes in time-series data stored either in a CSV file or on a graphite server. It is designed to be easily integrated into build pipelines \cite{CONTINUOUSBENCHMARKING} and provide automated performance analysis that can decide whether code should be deployed to production. As well as printing change point data on the command-line, Hunter also includes support for Slack and can be configured to send results to a Slack channel.

\subsection{Data Source}
Hunter can run analysis on data pulled from a graphite server or from data contained in an CSV file. Graphite support was necessary to integrate Hunter into our testing and deployment workflow. If developers are not using graphite as their central repository of benchmark data, the CSV support provides a common denominator for feeding data to Hunter.

\subsection{Configuration}
The data sources that Hunter uses are specified in a YAML configuration file. This configuration file has sections for graphite servers, Slack tokens, and data definitions. Hunter even supports templating which allows common definitions to be reused and avoids test definition duplication. Since we routinely use Hunter on hundreds of tests and metrics, the template feature helps to keep our configuration file small. For example, we use graphite metric prefixes to group related metrics together so that all metrics for a specific Apache Cassandra version are linked by a common string.

One example of this is the test db.20k-rw-ts.fixed, a benchmark running on Datastax Enterprise
that performs read and write operations at a fixed rate of throughput. We run this test in both
a configuration with replication factor 1 and with replication factor 3 and yet despite this
difference we can reuse around 95\% of the Hunter configuration because the metric types are the same.

Below is an example configuration file.

\begin{verbatim}
graphite:
  url: http://graphite.local
  suffixes:
    - ebdse_read.result

templates:
  common_metrics:
    metrics:
      throughput:
        scale: 1
        direction: 1
      p99:
        scale: 1.0e-6
        direction: -1

tests:
  db.20k-rw-ts.fixed:
    inherit:
    - common_metrics
    tags:
    - db.20k-rw-ts.fixed.1-bm2small-rf-1
    prefix: performance_regressions.db.20k-rw-ts.fixed

\end{verbatim}

For test data in CSV files, Hunter allows users to specify attributes of the file such as file path, which columns contain timestamps, which contains metrics, and the delimiter character used to separate fields on each line.

\subsection{Continuous Integration}
Since Hunter is a simple Python application, it has proven trivial to connect with different teams’ CI pipelines. We use a docker image to run Hunter against daily performance test results which are stored on a central graphite server. The Docker image is launched from a Jenkins job that runs once a day.

\subsection{Sending Results to Slack}
After running change point detection on a given time-series, Hunter can submit the results of its analysis to a Slack channel. We have found that this is the perfect location to notify developers of changes in performance mainly because each channel is already categorised by team or project. Developers usually triage the results of Hunter by investigating any unexpected changes in performance to identify whether there is a genuine change in performance for the product or the result was caused by noise in the workload or the platform.

Having Hunter’s results displayed in such a prominent location as Slack channels has resulted in improvements to the underlying infrastructure used to run performance tests. When test results show frequent fluctuations because of noise from the platform, one of our teams improved the stability of those platforms so that they are provided with more actionable results from Hunter.

\section{Implementation}
Hunter is built on top of the E-divisive Means algorithm available in the signal\_processing\_algorithms library from MongoDB \cite{MONGOSIG} but we have extended it in two ways to improve its efficiency (so that we can generate results faster) and to get repeatable results when performing multiple iterations on the same data set.

\subsection{E-divisive Means Algorithm}
The E-divisive means \cite{EDIV} is an offline change point detection algorithm that uses
hierarchical estimation to estimate the number and locations of change points in a distribution.
Since it’s a nonparametric method, it makes no assumptions about the underlying data distribution or
the change in distribution and is well suited for use with benchmarking data that is often
non-normal. The hierarchical aspect comes into play when deciding which collection of data points to
search for change points. E-divisive means divides the time-series into two parts and recursively
searches for change points.

Individual points are tested using a test statistic from previous change points which the literature
calls \^{q}, and the p-value of \^{q} is determined using random permutation testing which requires multiple calculations. Using random permutations comes with a performance cost and we found that detecting change points took an unreasonably long time for our data set. Additionally, because the permutations are random we found that the results of Hunter were non-deterministic and varied from run to run. It is possible to reduce the non-determinism in the results by increasing the number of permutations but this has the negative effect of increasing Hunter’s runtime. In our case running Hunter using the standard E-divisive means algorithm on hundreds of data points for a single test and single metric took 2-3 seconds. But to validate a nightly build or release, developers need the ability to run change point detection on tens of tests where each test recorded tens of metrics. This would push the runtime to several minutes, which was no longer ideal.

\subsection{Significance Testing}
When initially developing Hunter we profiled the code to understand which parts were taking the
longest to detect change points in our data. We discovered that the vast majority of the time was
spent performing significance testing. This wasn’t entirely surprising given the use of the \^{q} statistic and its reliance on random permutations. We switched to using Student’s t-test and saw the runtime of Hunter reduce by an order of magnitude as well as providing consistent results when run multiple times on the same data set. While Student’s t-test is not a robust measure of statistical significance for arbitrary data sets, it turned out it works extremely well for our scenario. 

We also tested using the Mann-Whitney U test. This would have been appealing since, unlike the
Student's t-test, it is a non-parametric test that doesn't assume the input data is normally
distributed. But it turned out to not behave very well on small amounts of data, as it requires ~30
points to be conclusive. In contrast both the original E-Divisive, and our Student's t-version, are
able to find changes in extremely short time series with only 4-7 points. Since E-Divisive is a hierarchical algorithm that splits the original time series into ever smaller windows, this is a significant difference.

\subsection{Fixed-Sized Windows}
As we began using Hunter on larger and larger data series, we discovered that change points identified in previous runs would suddenly disappear from Hunter’s results. This issue turned out to be caused by performance regressions that were fixed shortly after being introduced. This is a known issue with E-divisive means and is discussed in \cite{MONGOCPD}. Because E-divisive means divides the time series into two parts, most of the data points on either side of the split showed similar values. The algorithm therefore, by design, would treat the two nearby changes as a temporary anomaly, rather than a persistent change, and therefore filter it out. 

\begin{figure}
	\includegraphics[width=0.45\textwidth]{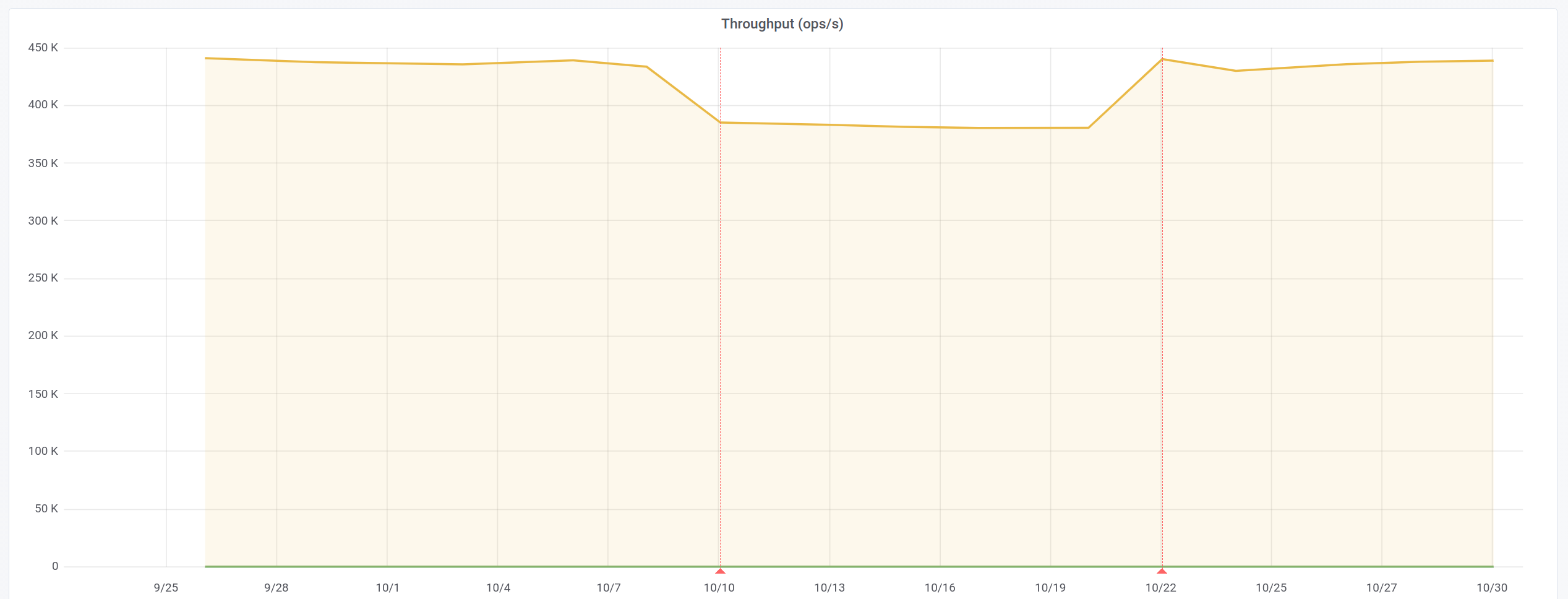}
	\caption{Temporary anomaly example}\label{Figure 1}
\end{figure}

Figure 1 illustrates this issue.

Our solution to this problem was to split the entire time series into fixed-sized windows and run the E-divisive means algorithm on each window individually. Change points that exist at window boundaries require special attention since change point detection algorithms in general are unable to identify whether the most recent point in a data series is a change point. To address this problem Hunter allows the windows to overlap and care is taken so that a change point isn’t reported multiple times because it exists in multiple windows.

\subsection{Weak Change Points}
Splitting the data series into windows partially addresses the problem of missing change points in
large data sets, but we also needed a method of forcing the E-divisive means algorithm to continue
recursively analysing the data series. The E-divisive means algorithm terminates when the
significance test, Student's t-test in Hunter, fails. If the algorithm first selects a change point with a p-value above the threshold set by the user (usually 0.05), it will terminate immediately, even if it would have detected change points below the p-value had it continued. We refer to change points that fail the significance test but would lead to other points below the p-value as weak change points.

The process of handling weak change points has two steps. First, we use a larger p-value threshold
when splitting so that it allows detection of weak change points. Second, we reevaluate the p-values and merge the split data series in a bottom-up way by removing change points that have a p-value above the smaller, user-specified threshold. We found that without forcing recursion to continue Hunter would miss some change points. Our modification results in much more accurate p-values.

Additionally, we filter out change points that show a small relative change, e.g. change points
where the difference in metric value is below 5\%. This relative threshold acts as a filter to discard change points that are not actionable, i.e. change points that are too small for developers to reproduce or verify a fix.

\section{Evaluation}
We evaluated our algorithms using benchmark data taken from a daily Gatling \cite{GATLING} performance test on Datastax Enterprise.
The benchmark data was saved to a CSV file and passed to Hunter using the
following command-line: \begin{verbatim}poetry run hunter analyse db-gatling.csv\end{verbatim}.

The data in db-gatling.csv contains 175 entries and covers 15 months’ worth of data. There are multiple performance changes contained within, both improvements (higher throughput or lower latency) and regressions (lower throughput or higher latency).

\begin{table*}
\caption{Performance and result accuracy for different significance tests in Hunter's e.divisive implementation.}
\centering
\begin{tabular}{|c | c | c | c | c|}
\hline
Algorithm & Mean Duration & Mean 95\% CI & Change points & Change points stddev\\
\hline
Permutation & 2.221 & 2.209, 2.233 & 16 & 1.174 \\
Student's t-test & 1.863 & 1.853, 1.873 & 20 & 0 \\
Student's + Weak Change Points & 1.594 & 1.584, 1.603 & 16 & 0 \\
\hline
\end{tabular}
\end{table*}

We opted for reading the data from a CSV file to avoid network communication delays with the
graphite server influencing the duration of each run. Every algorithm was run 30 times on the same
CSV file and the mean value, along with 95\% confidence intervals, are reported in Table 1.

Since we found that the permutation algorithm produced unstable results, we have also included the average number of change points detected for each of the algorithms in Table 1 as well as the standard deviations.

\subsection{Quickly Reverted Regressions}
Around 2020-10-10 on the graph in Figure 1 we can see a drop in the throughput. This performance regression was caused by a change to the way network packet decoding and processing was done in Datastax Enterprise. This problematic change was reverted on 2020-10-21 which explains why the throughput metric returns to previous values shortly after. Two red lines demarcate the data range where the regression is present. This is a known problem with change point detection and is explicitly mentioned in \cite{MONGOCPD}.

Both the Student's t-test and weak change points algorithms detected this regression and revert in each of the 30
runs through the data. The permutation based algorithm, only detected these changes for 15 of the 30
runs, or 50\% of the time.

\subsection{MongoDB Performance Test Result Dataset}

We also used the publicly available MongoDB Performance Test Result Dataset \cite{MONGODATA} to
compare the performance of E-divisive means with random permutations, and Student's t-test with 
weak change points filtering as the statistical significance test. This is the same data set used in
\cite{MONGOCPD}, so this analysis should be comparable and familiar to the emerging change point
detection community.

We arbitrarily selected 5 tests from the microbenchmark suite, and only focus on the
\emph{max\_ops\_per\_sec} result. All results are from task \emph{misc\_read\_commands} and 
variant \emph{linux-wt-standalone}. The 5 timeseries are shown in Figure 2. To avoid clutter, only one
timeseries was decorated with the change points found, but the results for the other 4 are similar.

As discussed above, the original E-divisive algorithm is not deterministic. Table 2 shows how many
change points were found for 100 iterations of each timeseries. The results are alarming. For example
for \emph{Remove.IntNonIdNoIndex} (row 5) it finds 4 change points 43\% of the time, 3 points 50\%
of the time, but 7\% of the time it finds zero change points!

Figure 2 also shows the other main issue that we have addressed. When a regresion is quickly followed
by a fix or rebound, then the original algorithm tends to ignore one or both changes as noise. The
9th change point in the graph is such an example, it is only found due to the approach with fixed sized
windows in this work.

Finally we can clearly see that our implementation finds many more change points. (The red diamonds
are found only with the Student's t-test configuration.) This is an expected
result as most of the modifications are motivated by making the algorithm as sensitive as possible.
Whether all 16 change points are meaningful is ultimately a subjective judgement, but looking closely
at the graph one can at least understand why the algorithm would have chosen each point. The implementation
offers to filter out changes that are too small to be actionable, but this feature was not used in
these tests, as we wanted to show the full output of our modified E-divisive algorithm, without post
filtering.

An obvious question we can already anticipate is that if the original implementation from MongoDB
performs thís poorly, how come MongoDB itself has used it so successfully? The answer is that a
a higher level in MongoDB's performance CI was designed such that if a point was once flagged as a
change point, the system will forever remember it. This way change points would not randomly disappear
while a developer is already working to fix it, and likewise a point marked as a false positive will
remain muted and not re-appear the next day. But the effect relative to the problems highlighted in
this paper is that the overall MongoDB CI system will eventually find and remember all change points,
even if on a given day the E-divisive algorithm may stop early and only return a subset. This higher
level system was documented in \cite{DSI} and the recorded change points are also part of \cite{MONGODATA}.

\begin{figure}
	\includegraphics[width=0.45\textwidth]{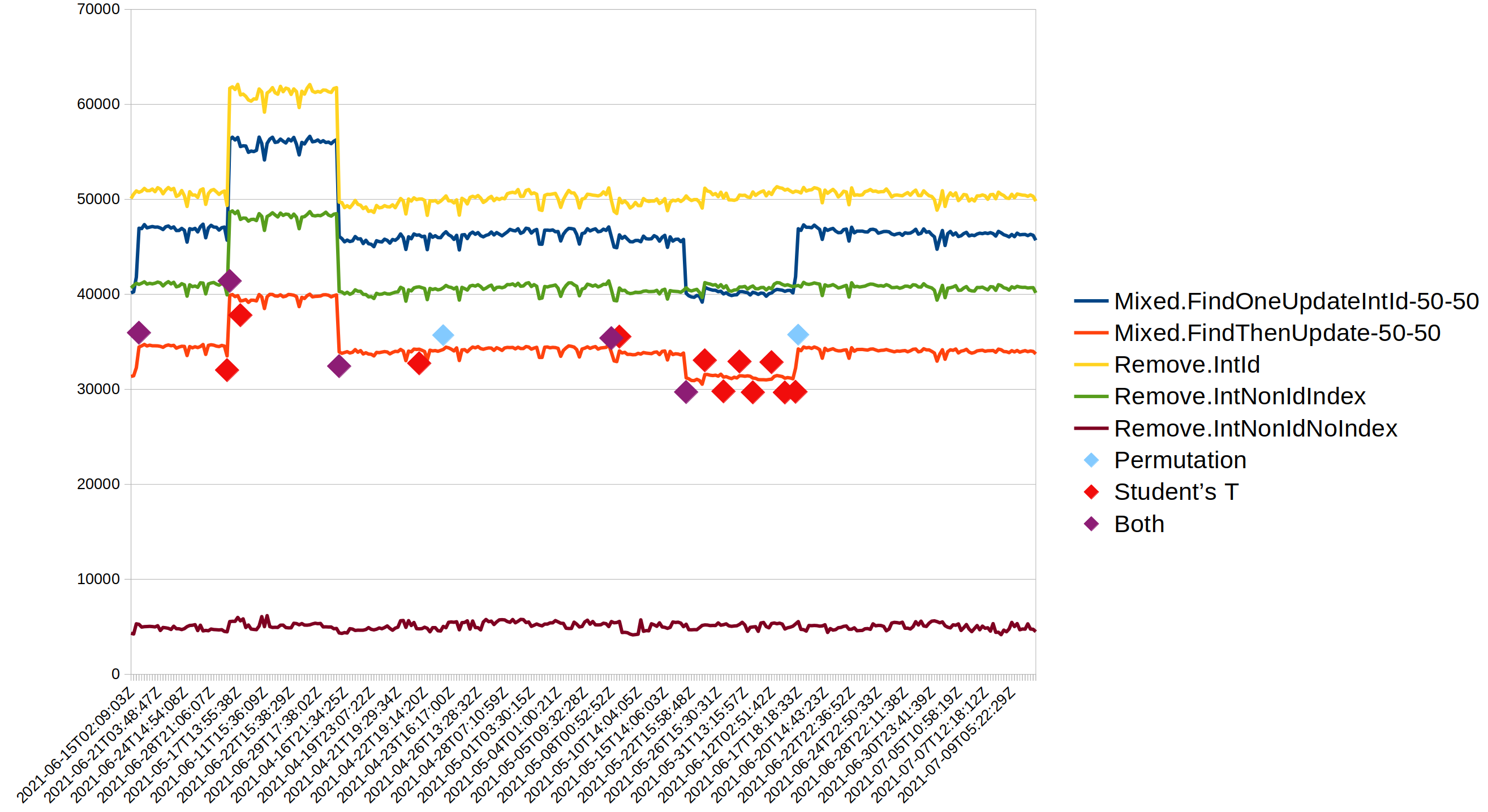}
	\caption{5 tests from the public MongoDB microbenchmarks dataset.}
\end{figure}

\begin{table*}
\caption{Distribution of nr of change points found with different statistical tests. (MongoDB data set, 100 iterations)}
\centering
\begin{tabular}{|c | c | c | c|c | c | c | c|c | c | c | c|c | c | c | c|}
\hline
Algorithm & Test name & 0 & 1 & 2 & 3 & 4 & 5 & 6 & 7 & 8 & 9 & 12 & 14 & 16 & 18 \\
\hline
Permutation & Mixed.FindOneUpdateIntId-50-50 & & & & & 1 & & & 1 & & 98 & & & &  \\
Permutation & Mixed.FindThenUpdate-50-50 & & & & & 1 & & & 99 & & & & & & \\
Permutation & Remove.IntId & & & & 64 & & 32 & 4 & & & & & & & \\
Permutation & Remove.IntNonIdIndex & & & & 61 & & 34 & 5 & & & & & & &\\
Permutation & Remove.IntNonIdNoIndex & 7 & & & 50 & 43 & & & & & & & & & \\
\hline
Student's + Weak & Mixed.FindOneUpdateIntId-50-50 & & & & & & & & & & & & & & 100 \\
Student's + Weak & Mixed.FindThenUpdate-50-50 & & & & & & & & & & & & & 100 & \\
Student's + Weak & Remove.IntId & & & & & & & & & & & & 100 & &  \\
Student's + Weak & Remove.IntNonIdIndex & & & & & & & & & & & & 100 & &  \\
Student's + Weak & Remove.IntNonIdNoIndex & & & & & & & & & & & 100 & & & \\
\hline
\end{tabular}
\end{table*}

\subsection{Evaluation with a Dataset with Known Change Points}

One weakness in the above analysis, and to our knowledge all previous literature published on this
topic so far, is that judging the accuracy of the algorithm or its implementation is always subjective.
Ultimately it's the human evaluator who decides whether a reported change point is a true positive,
or "useful", or "actionable". And note that those may not be the same! This is because an objective
truth about the correct set of change points is not available. If we had that knowledge, we would
not have needed this system to begin with.

It's of course possible to generate synthetic timeseries with changes injected at known steps, such
as a sine wave or even white noise, where the mean or amplitude is changing at discrete points.
However these tests tend to feel naive and E-divisive performs quite well against them.

To obtain a real data set, we employed Chaos Mesh\cite{CHAOS} to artificially generate network latency in the system
under test. In other words we artificially injected real changes, at known points in time, into a real
benchmark producing otherwise realistic results. The benchmark used was to test Cassandra with the 
same toolchain used for CI\cite{FALLOUT}.

In order to create different time series, we decided on a group of variables that we varied to
generate varying scenarios. We created 9 different scenarios by altering the values of the following
variables: number of changepoints, magnitude of change of variance between groups, magnitude of change
between groups and the length of groups. A group is defined as the set of points occurring between
two changepoints. Each scenario contains 5 test series, each with a minor variation. The scenarios
themselves can be grouped into three categories. The first scenario, change in mean, creates change
points by changing the mean of the groups. The variance remains relative constant. Similarly,
change in variance has constant mean and varying variance. This case is great to replicate noisy
environment's. Change in both mean and variance realistically replicates noisy environment's with random
latencies. 

Note that the timeseries used for this evaluation is different from those in previous sections.
Whereas previous timeseries have been a sequence of (nightly) builds, and the data points represent
values like average throughput during a test, in this evaluation the timeseries is from a single
benchmark, and the values are snapshots each second. This is because waiting for a year to create a
time series of true nightly builds was not practical.

\subsubsection{Evaluation Metrics}

Having obtained data sets with known change points, we can now employ objective statistical tests
to measure the accuracy of Hunter. Essentially we have recast the evaluation task as a machine
learning problem, where an algorithm is expected to produce a known output from a given input training
dataset.

We will evaluate hunter using two metrics, F1 score and Rand Index. In particular we will be
evaluating the \textbf{p99} metric. p99 indicates that 1 in every 100 users will encounter latency.
It is a common industry standard and also used for performance targets when developing Cassandra.

\subsubsection{True Positives}
We will be using the following two variables to represent the sets of ground truth and predicted points:

$$X^{*}: \text{set of ground truth}$$
$$X : \text{set of predicted points}$$

True positives are defined as the set of change-points in the detected class that are real change points i.e they are present in the set.
\textit{M} represents the scope of error. If the difference between the predicted points and the true changepoint in less that or equal to \textit{M} we will consider it as a correcty predicted changepoint.
$$TP(X , X^*) := \{x \in X | \exists  x^* \in X^*  s.t. |x- x^* | \leq M\}$$

It was important to ensure that there were no duplication's ie. if two points in the predicted set were in the margin of error of the same point in $X^{*}$. A  changepoint in $X^{*}$ was marked as \textit{visited} once a point in $X$ was within $M$ and added to $TP(X,X^{*})$ set and can not be considered again. 

\subsubsection{F1 Score}
The reasons for using the F1 Score to calculate the accuracy of hunter are that it is unaffected by the size and the density of data, it penalizes false positives and credits correct detections.

F1 score is defined as

$$F1 =2* \frac{precision * recall}{precision + recall}$$

Precision is the proportion of predicted change points that are true change points:\cite{SELECTREVIEW}
$$Precision  = \frac{|TP(X,X^*)|}{|X|^*}$$ 

Recall is the proportion of true change points that are well predicted:\cite{SELECTREVIEW}

$$Recall = \frac{|TP(X,X^*)|}{|X|}$$

\subsubsection{Rand Index}

Another metric we evaluated on was the rand index. 

$$RandIndex(X,X^*) = \frac{ TP + TN}{TP + TN+ FP+ FN}$$

\begin{itemize}
    \item \textbf{TP} : correctly predicts the positive class : True change points calculated
    \item \textbf{TN} : correctly predicts the negative class :  None in this case
    \item \textbf{FP} :  model incorrectly predicts positive class : $|X^*| - |TP(X,X^*)|$
    \item \textbf{FN} :model incorrectly predicts negative class: $|X| - |TP(X,X^*)|$
\end{itemize}

\subsubsection{Benchmarking against PELT and DYNP}

With an objective truth to benchmark against, it also becomes possible to compare Hunter against alternative change point detection algorithms. We therefore also present results from two other well-known offline algorithms PELT and DYNP. These algorithms were used using the ruptures package in python.\cite{SELECTREVIEW}

\subsubsection{Results}
We ran hunter over 45 test runs. The test runs were ran on a GKE cluster using Kubernetes. The cluster was ran on 4 x \textit{n2-standard-4} nodes in zone \textit{us-central1-a}. 

All the algorithms were evaluated on two metrics. It can be seen that hunter has consistently outperformed both \textit{pelt} and \textit{dynp} on both the metrics. In all the experiments we had an margin of error as \textbf{10 seconds}.

\begin{figure}[b]
	\includegraphics[width=0.45\textwidth]{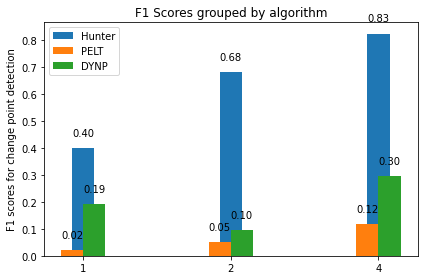}
	\caption{Correlation between F1 and number of points }\label{Figure 1}
\end{figure}

\subsubsection{Correlation to the number of points}

There is a positive correlation between the number of points and accuracy. As the number of points increase so does hunter's performance. For Figure 3 it can be seen that hunter outperformed both PELT and DYNP with a huge margine.

\subsubsection{Correlation between delta error and algorithms}
Hunter's performance increases if we allow a larger margin of error. Hunter is able to get an \textit{F1 score} of \textbf{0.1481} with an delta error of one second, where PELT and DYNP need a margin of error of at least 3 seconds to get a non-zero score. This is a key characteristic why E-divisive has served us well for detecting regressions in CI. Preferably we like to know the exact commit that caused a regression, not just the general area whereabouts a regression is suspected. E-divisive is superior in satisfying especially this requirement! The performance of all algorithms increases drastically as we give them slightly more flexibility in terms of margin of error. With a margin of error of 4 seconds we see that the performance increases to \textit{0.612} for Hunter. DYNP starts to catch up with Hunters accuracy at 15 seconds.

\begin{figure}[H]
	\includegraphics[width=0.45\textwidth]{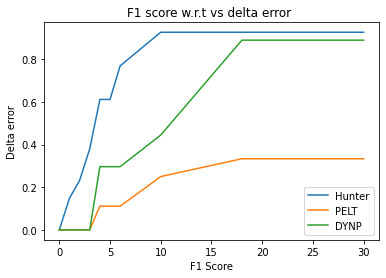}
	\caption{Correlation between F1 and delta error }
\end{figure}

\begin{table*}[!ht]
\caption{Evaluation results}
\centering
\begin{tabularx}{\textwidth}{
>{\hsize=0.4\hsize\bfseries\RaggedRight}X
*{6}{>{\hsize=0.1\hsize\centering\arraybackslash}X}}
    \toprule

\textbf{Algorithm}
    & \mcc{\textit{Hunter}} & \mcc{\textit{Pelt}} & \mcc{\textit{Dynp}} \\
    \cmidrule(lr){2-3} \cmidrule(lr){4-5} \cmidrule(lr){6-7}%
Metric     & $F1$ & $Rand$ & $F1$ & $Rand$ & $F1$ & $Rand$ \\
    \midrule
\multicolumn{7}{c}{\textbf{Single Change Point}} \\
    \midrule
 Scenario1   & 0.261904 & 0.166667 & 0.0 & 0.0 & 0.0 & 0.0 \\
    \addlinespace%
Scenario2    & 0.466667 & 0.305556 & 0.027027 & 0.013698 & 0.285714 & 0.166667\\
    \addlinespace%
Scenario3    & 0.466667 & 0.305556 & 0.040556 & 0.020698 & 0.285714 & 0.166667 \\
    \addlinespace
    \midrule
\multicolumn{7}{c}{\textbf{Two Change Points}} \\ \midrule[0.3pt]

 Scenario4   & 0.666667 & 0.666667 & 0.060606 & 0.031746 & 0.190476 & 0.111112 \\
    \addlinespace
 Scenario5   & 0.888889 & 0.833334 & 0.090909 & 0.047619 & 0.095238 & 0.055556 \\
    \addlinespace
 Scenario6   & 0.490476 & 0.327777 & 0.0 & 0.0 & 0.0 & 0.0 \\
    \addlinespace
\midrule
\multicolumn{7}{c}{\textbf{Four Change Points}} \\ \midrule[0.3pt]

 Scenario7   & 0.925926 & 0.866667 & 0.249999 & 0.142857 & 0.444445 & 0.285714 \\
    \addlinespace
 Scenario8   & 0.731313 & 0.579365 & 0.085713 & 0.045073 & 0.095238 & 0.055556 \\
    \addlinespace
 Scenario9   & 0.818182 & 0.714286 & 0.016460 & 0.00843 & 0.0 & 0.0 \\
    \addlinespace
    \bottomrule
\end{tabularx}
\end{table*}

\section{Lessons Learned}
We have now been operating Hunter for multiple teams for close to 2 years. In that time we’ve made a number of improvements in addition to the algorithmic changes covered in Section 3. The lessons we have learned, and the changes made in response, helped Hunter to become the de facto choice for statistical significance detection inside of DataStax.

\subsection{More Data Points Are Better}
We originally started off with 2 weeks worth of data points. Given that performance tests were run once a day this gave us 14 data points. This decision was primarily because we wanted to avoid the delay in collecting lots of data from our graphite server. This proved to be far too few data points to get meaningful results from Hunter and we increased it to a month (around 30) by default. This wider time range has allowed Hunter to deal with noise in the results much better and now we see fewer false positive change points. We plan to experiment with data sets covering a longer period of time in the future to see whether we can reduce the false positive rate even further.

\subsection{New Change Points Matter Most}
Once a change point has been reported to a developer it does not make sense to keep reporting it. When Hunter discovers many change points, reporting them via Slack can make the results overwhelming and make it difficult for developers to analyse. Things are made worse if a change point signals a performance regression that has since been fixed because Hunter will report both the old regression and more recent improvement as separate changes.

To quieten the output of Hunter’s Slack feature, we capped results to only show change points from the last 7 days. While this does ignore valuable data because the magnitude of the change point can be updated as new data is processed, those changes are not important enough to spam everyone on the Slack channel. In the case where developers need to see the full list of results they can run Hunter manually on the data series.

\subsection{Change Point Detection Cannot Fix Noisy Data}
One of the teams using Hunter was afflicted with frequent change point messages via the Slack bot.
After investigating these change points they discovered that the performance of the application
hadn’t changed, rather the change in benchmark results was caused by unstable hardware performance
in a private data center. Changes of +- 10\% for the median latency were typical.

While Hunter can detect statistically significant changes in time series data, it is still not impervious to data that contains wildly fluctuating points such as that produced by running benchmarks on untuned hardware.

However, the fact that the team was unable to fully take advantage of Hunter motivated them to investigate
the underlying issue and then migrate their benchmarks and tests to the cloud, which was shown to
produce more repeatable results than the internal benchmarking lab hardware. After the migration,
the benchmark results were much more stable and Hunter produced far fewer false positives. Figure 5 shows benchmark data for a single Paxos-based performance test. Before running the test on the public cloud on 2021-09-18 Hunter detected 3 change points per month, on average. All of these were false positives, that is changes in results that were not caused by software or configuration modifications. After the migration Hunter hasn’t detected a single false positive.

\begin{figure}
	\includegraphics[width=0.45\textwidth]{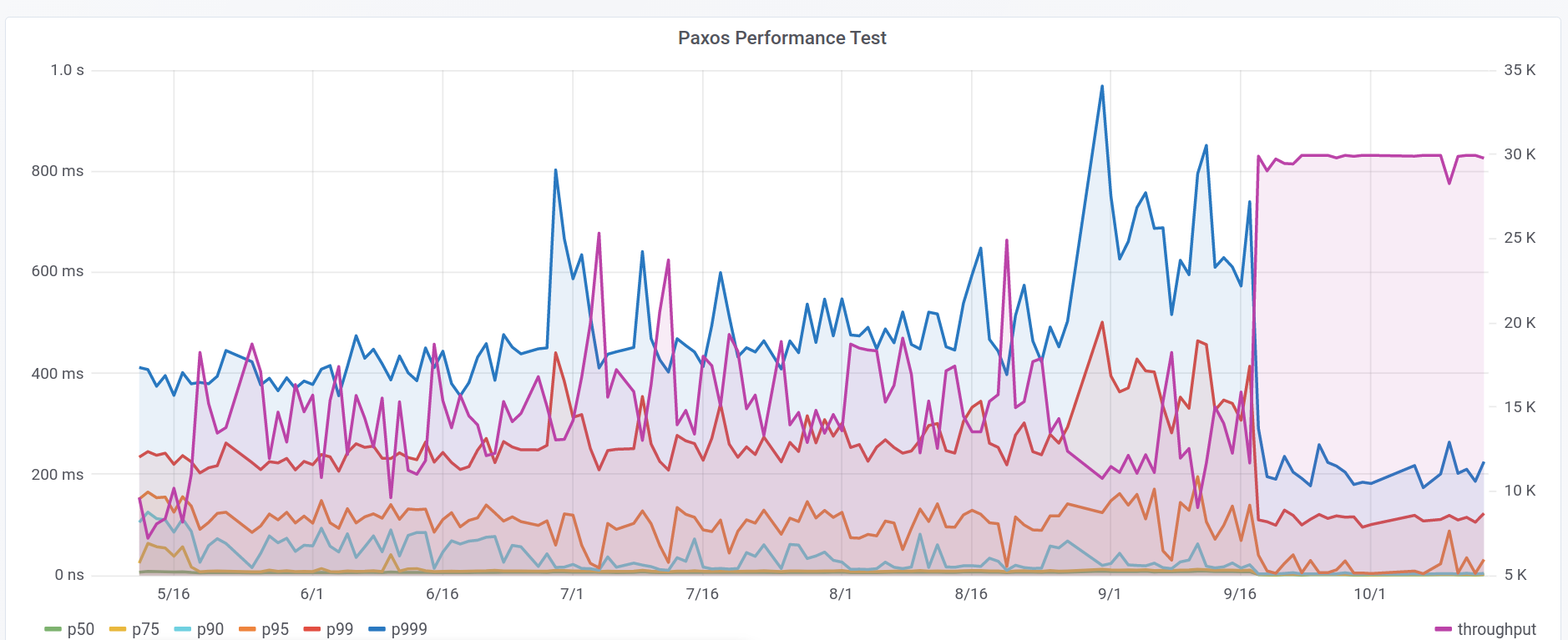}
	\caption{Unstable performance example}
\end{figure}

\section{Related Work}
We used the work in \cite{MONGOCPD} directly when creating Hunter and the novel contributions in this paper address some of the open questions posed there. Specifically, the authors of \cite{MONGOCPD} noted the bias inherent in the E-divisive means algorithm which favours detecting change points in the center of clusters. They make up for this bias by combining change point detection with anomaly detection which can identify large changes in performance as soon as the first data point in the new series is seen. Our use of windows for analysing data series addresses this same bias without resorting to anomaly detection which lacks the same sensitivity to changes as change point detection. Additionally, we are able to detect changes sooner, usually within 1-2 days.

Continuous Benchmarking \cite{CONTINUOUSBENCHMARKING} is a common technique for ensuring the performance of a product is maintained or improved as new code is merged into the source code repository and the literature includes examples of using change point detection \cite{MONGOCYCLE} and threshold-based methods to identify changes in software performance \cite{SAP} as part of a continuous integration pipeline. Multiple change point detection algorithms can also be combined into an ensemble which can outperform the individual algorithms \cite{MSTHESIS} when identifying performance changes.

The change point detection literature is vast and \cite{SURVEYCOOK} and \cite{SELECTREVIEW} provide excellent overviews and taxonomies of online and offline, supervised vs unsupervised, change point detection algorithms. In \cite{SURVEYCOOK} in particular, online sliding window algorithms are covered in detail.

Online change point detection has also been applied to identifying changes in performance. \cite{INTERVALCPD} combines change point detection with probabilistic model checking of interval Markov chains to promptly detect changes in the parameters of software systems and verify the system’s correctness, reliability, and performance.

Running performance tests in the cloud is known to be susceptible to performance variability \cite{BIGDATA} even when running the same software on the same hardware at different times. Historical performance data can be used to predict the future performance in cloud environments and \cite{CLOUDPREDICT} explores two change point detection algorithms, robseg and breakout, to predict variability in the cloud which enables users to plan repeatable experiments. \cite{CASESTUDYCPD} uses the E-divisive means algorithm to answer the question: does performance stability of serverless applications vary over time?

\section{Conclusion}
Detecting performance regressions across a range of product versions requires automation to be able to identify them quickly and without needing expert developers to manually detect them. Change point detection has emerged as a solution to this problem because of its ability to cope with noise in the data that is inherent to performance testing.

Hunter is an open source\cite{HUNTER} tool that uses change point detection to automatically identify changes in
time-series data, taken from either a graphite server or CSV file, and report the presence of change
points. Hunter extends the E-divisive Means algorithm to incorporate a Student's t-test which
removes the indeterminism present in the original version and provides reproducible results every time it is run on a single data series. We also introduced a sliding window technique to detect change points that are temporally close to each other. In addition to outperforming the original E-divisive means implementation, Hunter seems to also outperform two other well known algorithms, PELT and DYNP.

\section{Acknowledgments}
We are grateful to Guy Bolton King for his contributions to Hunter.

\bibliographystyle{ACM-Reference-Format}
\bibliography{hunter}

\end{document}